\documentclass[aps,prl,reprint,amsmath,amssymb,showpacs,superscriptaddress]{revtex4-1}
\usepackage{CJK}
\usepackage{graphicx}
\usepackage{dcolumn}
\usepackage{bm}

\usepackage{color}
\usepackage{hyperref}

\begin{document}
\begin{CJK*}{UTF8}{} 


\title{Nuclear energy density functionals from machine learning}

\author{X. H. Wu \CJKfamily{gbsn} (吴鑫辉)}
\affiliation{State Key Laboratory of Nuclear Physics and Technology, School of Physics, Peking University, Beijing 100871, China}

\author{Z. X. Ren \CJKfamily{gbsn} (任政学)}
\affiliation{State Key Laboratory of Nuclear Physics and Technology, School of Physics, Peking University, Beijing 100871, China}

\author{P. W. Zhao \CJKfamily{gbsn} (赵鹏巍)}
\email{pwzhao@pku.edu.cn}
\affiliation{State Key Laboratory of Nuclear Physics and Technology, School of Physics, Peking University, Beijing 100871, China}


\begin{abstract}
Machine learning is employed to build an energy density functional for self-bound nuclear systems for the first time.
By learning the kinetic energy as a functional of the nucleon density alone, a robust and accurate orbital-free density functional for nuclei is established.
Self-consistent calculations that bypass the Kohn-Sham equations provide
the ground-state densities, total energies, and root-mean-square radii with a high accuracy in comparison with the Kohn-Sham solutions.
No existing orbital-free density functional theory comes close to this performance for nuclei.
Therefore, it provides a new promising way for future developments of nuclear energy density functionals for the whole nuclear chart.
\end{abstract}


\maketitle

\end{CJK*}


Research on quantum mechanical many-body problems is essential in a wide variety of scientific fields.
By reducing the many-body problem formulated in terms of $N$-body wave functions to the one-body level with the density distributions, the density functional theory (DFT) has been enormously popular.
However, the quality of the DFT results crucially depends on the accuracy of the energy density functional, whose existence was proved by the Hohenberg-Kohn theorem~\cite{Hobenberg1964Phys.Rev.B864}, but the actual form is unknown and has to be determined with approximations.

Kohn and Sham first made DFT into a practical tool by computing the exact kinetic energy of an interacting many-body system with a noninteracting single determinantal wavefunction that gives rise to the same density~\cite{Kohn1965Phys.Rev.A1133}.
Since the kinetic energy is a major unknown part of the energy density functionals, the Kohn-Sham DFT is remarkably accurate~\cite{Kummel2008Rev.Mod.Phys.3}. Nevertheless, this approach involves not only the density but also the auxiliary one-body orbitals which need to be obtained by solving the so-called Kohn-Sham equation self-consistently.
This leads to enormous amount of work, in particular for systems with a large number of particles.

On the other hand, one of the aims of DFT is to express the energy solely as a functional of the density, i.e., orbital-free DFT. It is directly based on the Hohenberg-Kohn theorem~\cite{Hobenberg1964Phys.Rev.B864}, and is much more efficient than Kohn-Sham DFT due to the avoidance of the auxiliary one-body orbitals. However, it brings an inevitable trade-off between efficiency and accuracy. Instead of using orbitals to compute the kinetic energy, orbital-free DFT uses approximate kinetic energy density functionals, e.g., the well-known Thomas-Fermi functionals~\cite{Maruhn2010}, which renders it less accurate than Kohn-Sham DFT in most cases. Therefore, an accurate and computationally efficient orbital-free DFT is highly desired.

Of course, the stumbling block is that, as yet, no sufficiently accurate descriptions of kinetic energy with the density alone has been found. Despite more than seventy years of research and some tremendous progress, it is still a bottleneck of the DFT computations in various scientific fields. For nuclear physics, it is even more elusive because of the complicated short-range two nucleon interactions. In fact, for a local Skyrme energy density functional, the Thomas-Fermi solution of the ground-state density gives no spatial dependence whatsoever~\cite{Ring1980}. The description could be improved with higher-order corrections in the $\hbar$ expansion of the kinetic energy. However, there are still obvious deviations from the Kohn-Sham energy for the Thomas-Fermi approximation even corrected up to the $\hbar^4$ order due to the missing of sufficient quantum effects~\cite{Brack1985Phys.Rep., Centelles2007Ann.Phys.363}. As a result, most modern DFT calculations in nuclear physics are based on the Kohn-Sham approach~\cite{Bender2003Rev.Mod.Phys.121,Vretenar2005Phys.Rep.101,Meng2006Prog.Part.Nucl.Phys.470, Nakatsukasa2016Rev.Mod.Phys.45004}.

The recent rise in the popularity of machine learning (ML) has engendered many advances in various fields~\cite{Carleo2019Rev.Mod.Phys.45002,Carleo2017Science602}. Machine learning is a powerful tool for finding patterns in high-dimensional data, so it holds the promise of learning the energy density functional from sufficient ``energy-density data'' via induction. While applications of ML approaches to DFTs in condensed-matter physics and quantum chemistry have been proliferating in the past few years, the adoption in realistic computations is still in its infancy~\cite{Snyder2012Phys.Rev.Lett.253002,Brockherde2016Nat.Commun.872, Bogojeski2020Nat.Commun.5223,Meyer2020J.Chem.TheoryComput., Moreno2020Phys.Rev.Lett.,Li2021Phys.Rev.Lett.,Margraf2021Nat.Commun.}. For nuclear physics, ML applications can be traced back to early 1990s~\cite{Gazula1992Nucl.Phys.A1,Gernoth1993Phys.Lett.B1}, and recently, it has been more broadly adopted to nuclear masses~\cite{Utama2016Phys.Rev.C14311,Neufcourt2019Phys.Rev.Lett.62502,Niu2018Phys.Lett.B48, Neufcourt2018Phys.Rev.C34318,Wu2020Phys.Rev.C51301,Wu2021Phys.Lett.B}, charge radii~\cite{Utama2016J.Phys.G114002,Ma2020Phys.Rev.C14304,Wu2020Phys.Rev.C54323}, excited states~\cite{Lasseri2020Phys.Rev.Lett.162502, Bai2021Phys.Lett.B}, nuclear response functions~\cite{Raghavan2021Phys.Rev.C}, fission yields~\cite{Wang2019Phys.Rev.Lett.122501}, variational calculations~\cite{Keeble2020Phys.Lett.B135743, Adams2021Phys.Rev.Lett.}, extrapolations for many-body physics~\cite{Negoita2019Phys.Rev.C, Jiang2019Phys.Rev.C, Ismail2021Phys.Rev.C}, etc. However, the application of ML to develop density functionals for nuclear systems is still a blank area.

In contrast to the many-electron systems trapped in an external field in condensed-matter physics and quantum chemistry, nuclei are self-bound via strong and short-range nucleon-nucleon interactions. Thus, the nuclear DFTs are formulated in terms of the so-called ``intrinsic'' one-body density~\cite{Nakatsukasa2016Rev.Mod.Phys.45004, Bulgac2018Phys.Rev.C} (e.g., the density relative to the center-of mass). While ML is a balanced interpolation on known data and should be unreliable for densities far from the training set, fortunately, due to the saturation of nuclear forces, the shapes of the intrinsic nucleon density distributions for nuclei with different numbers of nucleons are roughly similar. In particular for heavy nuclei, the densities are roughly around the saturation density in the interior region, and decay exponentially in the surface region. This property is useful to simplify the ML model and/or reduce the demands of a huge training data set.

In this Letter, for the first time, ML is used to build an energy density functional for nuclei.
The employed ML model is kernel ridge regression (KRR)~\cite{Shawe-Taylor2004,Saunders1998KRR}, which is trained to describe the kinetic energy with the nucleon density alone.
An orbital-free density functional for nuclei is thus established after including the interaction energy.
As the first attempt, self-consistent calculations that bypass the Kohn-Sham equation are carried out for three nuclei $^{4}$He, $^{16}$O, and $^{40}$Ca, and they illustrate the key issues for applying ML to nuclear DFT problems.

The interaction energy $E_{\rm int}$ here is taken from the Skyrme functional SkP~\cite{Dobaczewski1984Nucl.Phys.A103}.
In contrast to the Kohn-Sham approach, the kinetic energy $E_{\rm kin}$ here is expressed solely as a functional of the density with the KRR approach~\cite{Shawe-Taylor2004,Saunders1998KRR},
\begin{equation}\label{KRR_finction}
  E_{\rm kin}^{\rm ML}[\rho(\bm{r})] = \sum_{i=1}^{m} \omega_{i} K(\rho_i,\rho).
\end{equation}
Here, $\omega_i$ are weights to be determined, $\rho_i({\bm r})$ are training densities, and $K$ is the kernel function, which measures the similarity between densities,
\begin{equation}\label{Kernel}
  K(\rho,\rho') = \exp\left[ -||\rho(\bm{r})-\rho'(\bm{r})||^2/(2A A'\sigma^2) \right].
\end{equation}
Here, $\sigma$ is a hyperparameter defining the length scale on the distance that the kernel affects, and the distance between two densities $||\rho(\bm{r})-\rho'(\bm{r})||$ can be calculated by vectorizing the densities on a series of discrete grids.
The factors $A$ and $A'$ are the nucleon numbers of the densities $\rho$ and $\rho'$, respectively.
They are introduced to scale the density distance, so that the kernel function $K$ can be directly applied to densities with different nucleon numbers.
This is important for applying the present ML approach to the whole chart of nuclei in the future.

The weights $w_i$ are determined by minimizing the loss function
\begin{equation}\label{Loss}
  L({\bm \omega}) = \sum_{i=1}^m \left( E_{\rm kin}^{\rm ML}[\rho_i] - E_{\rm kin}[\rho_i]\right)^2 + \lambda||{\bm \omega}||^2,
\end{equation}
where ${\bm \omega} = (w_1,...,w_m)$.
The second term with the hyperparameter $\lambda$ is a regularizer that penalizes large weights to reduce the risk of overfitting.
Minimizing the loss function yields
\begin{equation}\label{weight}
  {\bm \omega} = ({\bm K}+\lambda{\bm I})^{-1}{\bm E}_{\rm kin},
\end{equation}
where ${\bm K}$ is the kernel matrix with elements ${\bm K}_{ij} = K(\rho_i,\rho_j)$, ${\bm I}$ is the identity matrix, and ${\bm E}_{\rm kin}$ are the exact kinetic energies to be learned, i.e., $(E_{\rm kin}[\rho_1], \ldots, E_{\rm kin}[\rho_m])$.

The nuclear ground state is obtained by a variation of the energy density functional with respect to the density.
The variation of the interaction energy can be readily obtained, while for the kinetic energy, it reads
\begin{equation}\label{functional_derivative}
  \frac{\delta E^{\rm ML}_{\rm kin}[\rho]}{\delta \rho} =\sum_{i=1}^{m}
  \frac{w_i}{AA_i\sigma^2 }(\rho_i - \rho)K(\rho_i,\rho).
\end{equation}
It causes huge errors because Eq.~(\ref{functional_derivative}) does not contain information on how the energies change along all directions, but only directions in which it has training data.
There have been various ways to overcome this problem in the literature, see e.g., Refs~\cite{Snyder2015Int.J.QuantumChem.1102,Brockherde2016Nat.Commun.872}.
Here, we employ a prime local principal component analysis~\cite{Snyder2015Int.J.QuantumChem.1102,Abdi2010WIREsComputationalStatistics433}, and together with the newly proposed adaptive functional derivative and density renormalization recipes, the accuracy is found to be rather good for nuclei.
Once the functional derivative is obtained, the self-consistent ground-state density can be calculated with the gradient descent method starting from a trial density.
More details can be seen in the supplementary materials~\cite{[See Supplementary Material for detailed information on the generation of the machine learning data sets\rm{,} the illustration of prediction ability\rm{,} and the numerical techniques on the self-consistent iterations for the ground state] supplement1}.

In this work, we only consider spherical nuclei with the same proton and neutron numbers. The densities to be learned by the KRR network can be reduced to the one-dimensional radial densities with the metric $\tilde \rho(r) = 4\pi r^2\rho(r)$, and they are expressed in 501 discretized spatial mesh points from 0 to 20 fm.

The training data of kinetic energies and densities are prepared by solving the Schr\"odinger equation~\cite{Killingbeck1987J.Phys.AMath.Gen.} with a mean potential for noninteracting systems.
A combined Gaussian potential (CGP) is proposed to simulate the mean potential for nuclei, which is similar to the well-known Woods-Saxon potential for medium and heavy nuclei, but is more flexible and proper for light ones.
In the present work, a series of CGPs are generated to build the training, validation, and test sets of kinetic energies and densities with some physical constraints on the potential depth and the root-mean-square (rms) radius. The details can be seen in Ref.~\cite{[See Supplementary Material for detailed information on the generation of the machine learning data sets\rm{,} the illustration of prediction ability\rm{,} and the numerical techniques on the self-consistent iterations for the ground state] supplement1}. It should be noted that the training data of both kinetic energy and density can be calculated simply by solving single-particle Schr\"odinger equations with given mean potentials. This process is much easier than the Kohn-Sham calculation, where a self-consistent solution of the Kohn-Sham potential must be achieved.
\begin{figure}[!htbp]
\centering
\includegraphics[width=8.5cm]{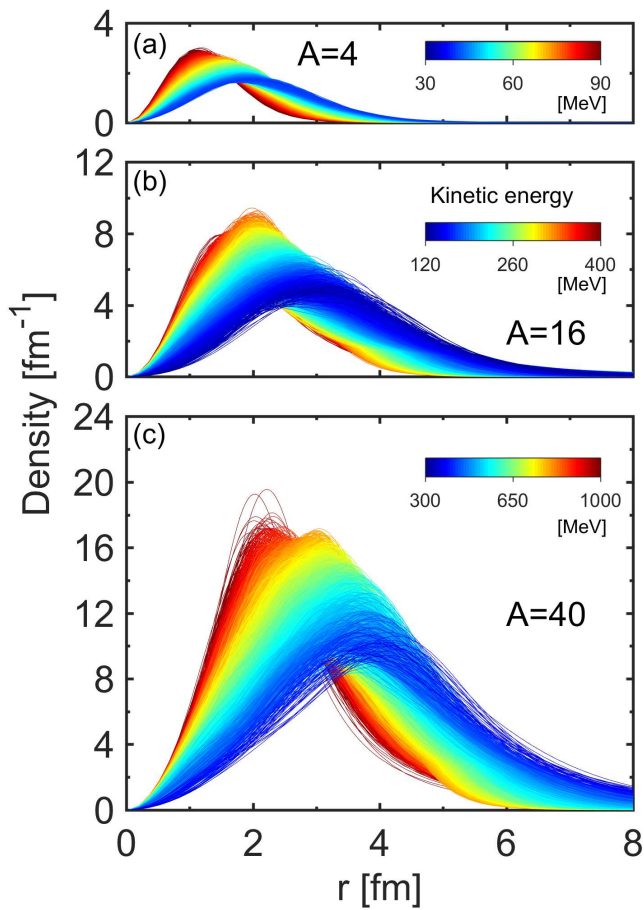}
\caption{(Color online). Training density distributions and the corresponding kinetic energies (color map) for nuclei with the total nucleon number $A=4$ (top), 16 (middle), and 40 (bottom).}
\label{fig1}
\end{figure}

In Fig.~\ref{fig1}, 30000 samples of the training density distributions and the corresponding kinetic energies for nuclei with the total nucleon number $A=4$, 16, and 40 (10000 ones for each nucleus) are depicted.
The lines give the density distributions and the colors represent the kinetic energies.
While the training kinetic energies range from several tens MeV to more than one thousand MeV for different systems, the kinetic energies per nucleon are roughly several tens MeV, which is quite reasonable for most nuclei.
The density distributions are closely related to the magnitudes of the kinetic energies according to the uncertainty principle, i.e., the tighter the density distribution, the higher the kinetic energy.

These 30000 training energy-density data are used to train the KRR network~(\ref{KRR_finction}), and the solution can be obtained via Eq.~(\ref{weight}).
Here, the hyperparameters $\lambda$ and $\sigma$ are determined by optimizing the ML performance on the validation set containing 3000 samples of data (1000 ones for each nucleus).
The resultant hyperparameters are $\lambda = 1.8\times10^{-12} $ and $\sigma = 0.58$ fm$^{-1}$.
Finally, a test set containing 3000 samples of data is used to provide an unbiased evaluation of the KRR training.

\begin{figure}[!htbp]
\centering
\includegraphics[width=8.5cm]{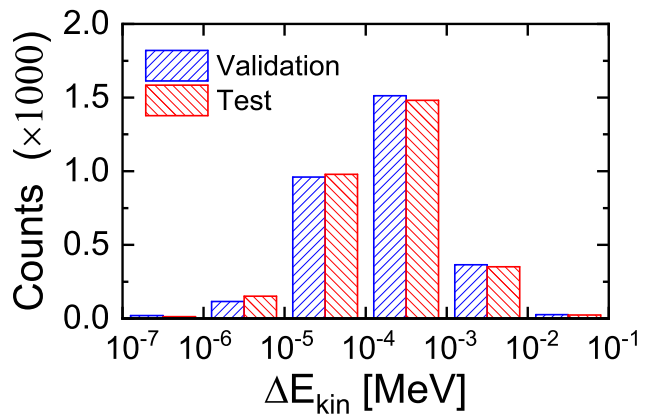}
\caption{(Color online). Statistical histogram of the deviations $\Delta E_{\rm kin}$ between the KRR predicted kinetic energies and the exact ones in the validation and test sets.}
\label{fig2}
\end{figure}

In Fig.~\ref{fig2}, the performance of the KRR functional is illustrated with the statistical histogram of the deviations $\Delta E_{\rm kin}$ between the KRR predicted kinetic energies and the exact ones in the validation and test sets.
It is seen that the KRR predictions nicely reproduce the kinetic energies in both validation and test sets and, in most cases, the deviations are below 1 keV, which is a very high accuracy for nuclear physics.
More importantly, the count distributions of the deviations $\Delta E_{\rm kin}$ are very similar for the validation and test sets.
This means that the performances of the present high-accuracy KRR functional are equally well for the two datasets and, thus, the functional is very good in its generalization ability.

\begin{figure}[!htbp]
\centering
\includegraphics[width=8.5cm]{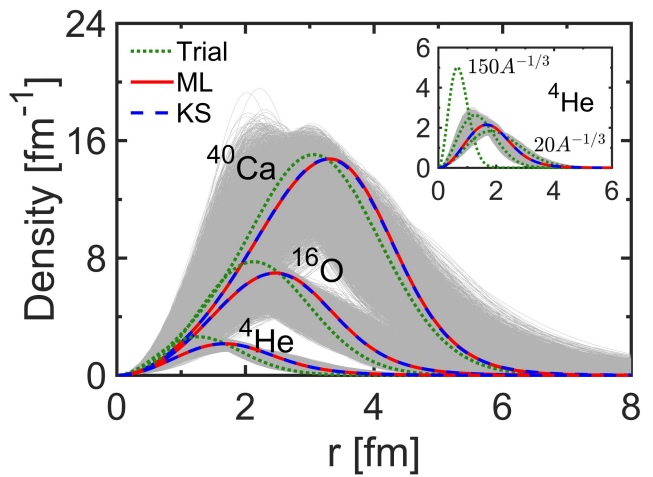}
\caption{(Color online). The trial (dotted lines) and self-consistent densities (solid lines) from the machine learning approach in comparison with the self-consistent densities obtained with the Kohn-Sham method (dashed lines) for $^{4}$He, $^{16}$O, and $^{40}$Ca.
The shaded region shows the extent of variation of the training densities.
The inset zooms in the densities for $^{4}$He with two additional trial densities (see text).}
\label{fig3}
\end{figure}

While the ML functional predicts the energy in a very high accuracy, for orbital-free DFT, one has to finally test the self-consistent procedures to find the density that minimizes the total energy.
The iteration starts from a trial density which is intuitively taken as the empirical nuclear density in a harmonic oscillator (HO) potential with $\hbar\omega = 41 \cdot A^{-1/3}$ MeV~\cite{Ring1980}.

Moreover, for self-bound nuclear systems, one has to consider the center of mass (c.m.) correction energy due to the translational symmetry breaking. There have been several ways to do this in the literature~\cite{Bender2000Eur.Phys.J.A,Zhao2009Chin.Phys.Lett.}, while one should in principle, calculate the c.m. corrections in the same way that was employed in the fitting of the adopted density functional.
For the SkP functional~\cite{Dobaczewski1984Nucl.Phys.A103} here, the c.m. correction energy is calculated with the diagonal terms of the microscopic corrections $E_{\rm c.m.}^{\rm mic}=-\frac{1}{2mA}\langle \hat{\bm P}^2_{\rm c.m.} \rangle$ before variation, i.e., $E_{\rm c.m.}^{\rm dic}= -\frac{1}{A}E_{\rm kin}$~\cite{Bender2000Eur.Phys.J.A}.
Note that a full consideration of $E_{\rm c.m.}^{\rm mic}$ is also possible by building a ML density functional for the c.m. correction energy via a similar training process for the kinetic functional.

In Fig.~\ref{fig3}, the trial densities and the obtained self-consistent ones are depicted in comparison with the Kohn-Sham ground-state densities for $^{4}$He, $^{16}$O, and $^{40}$Ca.
One can see that for all the three nuclei, the ground-state densities obtained with the present ML approach are in a very nice agreement with the Kohn-Sham ones.
The differences are less than 0.1 ${\rm fm}^{-1}$ for $^{40}$Ca, 0.04 ${\rm fm}^{-1}$ for $^{16}$O, and 0.01 ${\rm fm}^{-1}$ for $^{4}$He, respectively.
This clearly demonstrates the success of the present ML density functional for nuclei.
One may doubt that the trial densities here are coincidently too close to the self-consistent ones.
As seen in the inset of Fig.~\ref{fig3}, for $^{4}$He, the trial density  can actually be widely changed by tuning the HO strengths $\hbar\omega$ from $20\cdot A^{-1/3}$ MeV to $150\cdot A^{-1/3}$ MeV.
It is surprisingly found that even if the trial density is out of the training data region, the convergence is still stable and accurate.
The same conclusion holds true for $^{16}$O and $^{40}$Ca as well.
Therefore, one can conclude that the performance of the present ML density functional in finding self-consistent densities is quite robust.

\begin{table}[h]
\caption{Total energies $E_{\rm tot}$ (MeV), kinetic energies $E_{\rm kin}$ (MeV), and root-mean-square radii $\langle r^2 \rangle$ (fm) for $^{4}$He, $^{16}$O, and $^{40}$Ca obtained by the self-consistent Kohn-Sham and machine-learning approaches, in comparison with the data available~\cite{Tuli2011}. }
\begin{center}
\setlength{\tabcolsep}{1.4mm}{
\begin{tabular}{l l r r r}
\hline
\hline
 & & Kohn-Sham & Machine-Learning & Experiment \\
\hline
 & $E_{\rm tot}$             & -26.3700  & -26.3931  (0.0012)  & -28.2957    \\
 $^{4}$He & $E_{\rm kin}$    & 35.2138   & 35.2044   (0.0056)  & /           \\
 & $\langle r^2 \rangle$     & 2.1626    & 2.1628      (0.0002)  & 1.6755      \\
\hline
 & $E_{\rm tot}$             & -127.3781 & -127.1622 (0.1584)  & -127.6193   \\
 $^{16}$O & $E_{\rm kin}$    & 219.2875  & 218.3458  (0.6882)  & /           \\
 & $\langle r^2 \rangle$     & 2.8077    & 2.8113      (0.0047)  & 2.6991      \\
\hline
 & $E_{\rm tot}$             & -342.0645 & -341.8027 (0.5724)  & -342.0521   \\
 $^{40}$Ca & $E_{\rm kin}$   & 643.1100  & 642.9145  (1.6875)  & /           \\
 & $\langle r^2 \rangle$     & 3.4677    & 3.4652      (0.0055)  & 3.4776      \\
\hline
\hline
\end{tabular}}
\end{center}
\label{tab1}
\end{table}

To qualitatively evaluate the influence of the trial density on the self-consistent results, we generate 100 trial densities by randomly sampling the HO strengths $\hbar\omega$ between $20\cdot A^{-1/3}$ MeV and $150\cdot A^{-1/3}$ MeV for nuclei $^{4}$He, $^{16}$O, and $^{40}$Ca, respectively.
Starting from these trial densities, the self-consistent iteration produces a range of similar ground-state densities and energies.
Thus, the mean and standard deviation are taken as the final result and the corresponding uncertainty, respectively.
The obtained self-consistent kinetic energies, total energies, and the rms radii are compared with the Kohn-Sham results and the available data in Table~\ref{tab1}.
To compare with the data, the Coulomb energies are included with the Slater approximation~\cite{Beiner1975Nucl.Phys.A,Slater1951Phys.Rev.}.

One can see that the central values of the present ML results agree nicely with the Kohn-Sham ones within a rather small deviation below 0.5\% for all nuclei.
No existing orbital-free DFT comes close to this performance for nuclei.
The uncertainties for the total energies are smaller than those for the kinetic energies.
This is due to the fact that nuclei are self-bound systems with short-range repulsive and long-range attractive nucleon-nucleon interactions.
As a result, the kinetic and the interaction energies are elegantly balanced with the variation of the density around the equilibrium.

The present study is the headmost work to build nuclear energy density functional from ML, and it aims for a proof of principle that ML approach can be successfully employed to build nuclear orbital-free DFTs with high precisions.
For this purpose, only three nuclei are considered.
However, it has been checked that the prediction ability of the present ML density functional to systems with other mass numbers is quite good.
A detailed discussion can be found in Ref.~\cite{[See Supplementary Material for detailed information on the generation of the machine learning data sets\rm{,} the illustration of prediction ability\rm{,} and the numerical techniques on the self-consistent iterations for the ground state] supplement1}. Note that the computational complexity of the present KRR approach grows fast with the increase of the training set. Alternative ML approaches like neural network may be employed in the future, and some learning techniques (e.g., minibatch learning) can help to optimize the learning procedure.

In summary, for the first time, machine learning has been used to build an energy density functional for self-bound nuclear systems.
The kernel ridge regression is employed to learn the kinetic energy as a functional of the nucleon density alone, and together with the interaction energy, it provides a robust and accurate orbital-free density functional for nuclei.
Self-consistent calculations have been carried out for $^{4}$He, $^{16}$O, and $^{40}$Ca without solving the Kohn-Sham equations, and the obtained ground-state properties have been compared with the Kohn-Sham results.
It is found that the obtained self-consistent densities, energies, and radii for the ground states agree quite well with the Kohn-Sham ones.
The relative deviations are below 0.5\% for all nuclei.
No existing orbital-free DFT comes close to this performance for nuclei.
Therefore, it provides a new promising way for future developments of nuclear energy density functionals.

The pairing correlations could also be included in the present ML approach straightforwardly by treating the kinetic energy as a functional of normal and pairing densities.
This would massively reduce the complexity of three-dimensional (3D) large-scale DFT calculations for nuclei by bypassing the 3D Hartree-Fock-Bogoliubov (HFB) equation which is extremely expensive in computation for heavy nuclei.
For example, as mentioned in Ref.~\cite{Jin2017Phys.Rev.C}, even the most advanced HFB solver in 3D lattice space requires around 9300 charged core hours at the Titan supercomputers in Oak Ridge National Laboratory. For the most commonly used traditional HFB solver by the diagonalization procedure, the computational cost would be around 46000 CPU hours.
By bypassing the Kohn-Sham equations, the present ML orbital-free DFT could be much more efficient than the conventional Kohn-Sham approach, since its computational cost scales as $O(N)$ with the size of the system $N$, while for Kohn-Sham DFT, it scales as $O(N^3)$. This advantage holds for both static and time-dependent DFT calculations.
In fact, the time-dependent orbital-free DFT has been developed and applied to quantum many-body problems in many other fields, see e.g., Refs.~\cite{Domps1998Phys.Rev.Lett.5520, Ding2018Phys.Rev.Lett.145001, Covington2021Phys.Rev.B075119}.
It has shown a quite higher efficiency compared with the conventional time-dependent Kohn-Sham DFT calculations, which are rather time-consuming for nuclear many-body problems~\cite{Hashimoto2016Phys.Rev.C014610}.
Therefore, the present machine-learning approach provides a promising way to develop time-dependent orbital-free DFTs for nuclear systems, which will be beneficial for solving the complex nuclear dynamical problems, such as fission and fusion problems.

\begin{acknowledgments}
This work was partly supported by the National Key R\&D Program of China (Contracts No. 2018YFA0404400 and No. 2017YFE0116700), the National Natural Science Foundation of China (Grants No. 11875075, No. 11935003, No. 11975031, No. 12141501 and No. 12070131001), and the China Postdoctoral Science Foundation under Grant No. 2021M700256 and 2020M670013.
\end{acknowledgments}

\end{document}


\begin{CJK*}{UTF8}{} 


\title{Supplementary materials}

\author{X. H. Wu \CJKfamily{gbsn} (吴鑫辉)}
\affiliation{State Key Laboratory of Nuclear Physics and Technology, School of Physics, Peking University, Beijing 100871, China}

\author{Z. X. Ren \CJKfamily{gbsn} (任政学)}
\affiliation{State Key Laboratory of Nuclear Physics and Technology, School of Physics, Peking University, Beijing 100871, China}

\author{P. W. Zhao \CJKfamily{gbsn} (赵鹏巍)}
\email{pwzhao@pku.edu.cn}
\affiliation{State Key Laboratory of Nuclear Physics and Technology, School of Physics, Peking University, Beijing 100871, China}

\maketitle
\end{CJK*}



\subsection{Solution of radial Schr\"odinger equation}

In order to build the data sets of kinetic energies and densities, the radial Schr\"odinger equation with a mean potential $v(r)$ is solved numerically,
\begin{equation}
  \left[-\frac{1}{2m_N}\frac{\partial^2}{\partial r^2}+v(r)+\frac{l(l+1)}{2m_Nr^2}\right]\psi_{n_rl}(r)=\varepsilon_{n_rl}\psi_{n_rl}(r),
\end{equation}
where the nucleon mass is taken as $m_N = 939$ MeV, $\psi_{n_rl}(r)$ is the radial wavefunction labeled by the radial quantum number $n_r$ and the orbital quantum number $l$, and $\varepsilon_{n_rl}$ is the single-particle energy.

A combined Gaussian potential (CGP) is proposed to simulate the nuclear mean potential,
\begin{eqnarray}\label{potential}
  v(r) &=&  -a_1\exp\left[-\frac{(r-b_1)^2}{2c_1^2}\right] - a_2\exp\left[-\frac{(r-b_2)^2}{2c_2^2}\right] \nonumber \\
  &&-\frac{a_1(r-b_1)r}{c_1^2}\exp\left[-\frac{(r-b_1)^2}{2c_1^2}\right].
\end{eqnarray}
It is a linear combination of two Gaussian potentials with different depths ($a_1$ and $a_2$), centers ($b_1$ and $b_2$), and widths ($c_1$ and $c_2$), and the third term with no additional parameter is included to assure a reasonable asymptotic behavior of the derivative at $r=0$.

The shooting method~\cite{Killingbeck1987JPA} is used to solve the radial Schr\"odinger equation. The radial coordinate $r$ is discretized on a uniform grid of 501 points from 0 to 20 fm with spacing $\Delta r = 0.04$~fm, i.e.,
\begin{equation}
  r_i = i\Delta r,~~i=0,1,\cdots,500.
\end{equation}
To avoid the singularity of the centrifugal potential $\dfrac{l(l+1)}{2m_Nr^2}$ at the origin, the value of $r_0$ is taken as small as $\Delta r/10000$.
The boundary conditions for $\psi_{n_rl}(r)$ are taken as,
\begin{equation}
  \psi_{n_rl}(r_0) = \psi_{n_rl}(r_{500}) = 0.
\end{equation}

Then, for a given orbital angular momentum $l$, one can either calculate the wavefunctions starting from the ``left'' $\psi^{(l)}_{n_rl}(r_0) = 0$ and $\psi^{(l)}_{n_rl}(r_1) = 1$ to the ``right'' $\psi^{(l)}_{n_rl}(r_i)$ via
\begin{eqnarray}
\psi^{(l)}_{n_rl}(r_{i+1}) &=&
\frac{24-10\Delta r^2f(r_i)}{12+\Delta r^2f(r_{i+1})}  \psi^{(l)}_{n_rl}(r_i) \nonumber \\
&&-\frac{12+\Delta r^2f(r_{i-1})}{12+\Delta r^2f(r_{i+1})} \psi^{(l)}_{n_rl}(r_{i-1}),
\end{eqnarray}
with $f(r)=2m_N\left[\varepsilon_{n_rl} - v(r) - \dfrac{l(l+1)}{2m_Nr^2}\right]$, or from the ``right'' $\psi^{(r)}_{n_rl}(r_{500}) = 0$ and $\psi^{(r)}_{n_rl}(r_{499}) = 1$ to the ``left'' $\psi^{(r)}_{n_rl}(r_i)$,
\begin{eqnarray}
\psi^{(r)}_{n_rl}(r_{i-1}) &=&
\frac{24-10\Delta r^2f(r_i)}{12+\Delta r^2f(r_{i-1})}  \psi^{(r)}_{n_rl}(r_i) \nonumber \\
&&-\frac{12+\Delta r^2f(r_{i+1})}{12+\Delta r^2f(r_{i-1})} \psi^{(r)}_{n_rl}(r_{i+1}).
\end{eqnarray}
The single-particle energy $\varepsilon_{n_rl}$ can thus be determined by solving the equation at a match point $m_0$, e.g., $m_0 = 50$,
\begin{equation}
  {\rm Det}(\varepsilon_{n_rl})\equiv
  \left|
    \begin{matrix}
      \psi^{(r)}_{n_rl}(r_{m_0-1})  &\psi^{(r)}_{n_rl}(r_{m_0})\\
      \psi^{(l)}_{n_rl}(r_{m_0-1})  &\psi^{(l)}_{n_rl}(r_{m_0})\\
    \end{matrix}
  \right|=0.
\end{equation}
In practice, this equation is solved by the bisection search for the single-particle energy $\varepsilon_{n_rl}$.
The obtained eigenfunction is then normalized and further used to calculate various quantities.

For nuclei with $N=Z$, the density, kinetic energy, and radius are respectively calculated with
\begin{subequations}
\begin{align}
  \rho(r_i) = & 2\sum_{n_rl}f_{n_rl}\left[\psi_{n_rl}(r_i)/r_i\right]^2, \label{density}\\
  E_{\rm kin} = & 2\sum_{n_rl}f_{n_rl}\varepsilon_{n_rl}-\Delta r\sum_{i=0}^{500}4\pi r_i^2\rho(r_i)v(r_i),\label{kinetic} \\
  R_{\rm rms} = & \sqrt{\Delta r\sum_{i=0}^{500}4\pi r_i^4\rho(r_i)/A} \label{radius},
\end{align}
\end{subequations}
where $A$ is the mass number, $f_{n_rl}$ is the occupation number determined by equal-filling approximation.

\subsection{Data generation}

The six parameters $a_1$, $b_1$, $c_1$, $a_2$, $b_2$, and $c_2$ in the mean potential are randomly sampled in the range of $20<a_1<320$~MeV, $-5.0<b_1<0.5$~fm, $1.0<c_1<2.5$~fm, $-10<a_2<30$~MeV, $0.5<b_2<4.5$~fm, and~$0.3<c_2<1.0$~fm, respectively.
In addition, two physical constraints are employed to exclude unphysical densities and kinetic energies.
First, the potential should not be deeper than $-80$~MeV. Second, the calculated root-mean-square radius $R_{\rm rms}$ of nuclei  should be in the range of $0.9A^{1/3}$ fm and $1.5A^{1/3}$ fm, which is an empirical estimate of nuclear radii according to the basic properties of nuclear force.

We generate, in total, 36000 pairs of densities and kinetic energies for systems with $A = 4$, 16, and 40 (12000 ones for each system).
They are divided into three sets, i.e., 30000 pairs for the training set, 3000 ones for the validation set, and 3000 ones for the test set. In each set, the numbers of the data for systems with $A = 4$, 16, and 40 are equal.

\subsection{Prediction ability}

This part aims to illustrate the prediction ability of the machine-learning (ML) functional to systems that are not used in the training set.

We generate a training set, which includes 800 pairs of densities and kinetic energies for eight $N=Z$ systems with $A=16$, 26, 36, 46, 56, 66, 76, and 86 (100 pairs for each system).
The training data of kinetic energies and densities are obtained, as mentioned above, by solving the Schr\"odinger equations for noninteracting systems in randomly generated mean potentials. The six parameters $a_1$, $b_1$, $c_1$, $a_2$, $b_2$, and $c_2$ in the mean potential are randomly sampled in a relatively larger range to obtain proper densities for nuclei heavier than $A=40$.

\begin{figure}[!htbp]
\centering
\includegraphics[width=8.5cm]{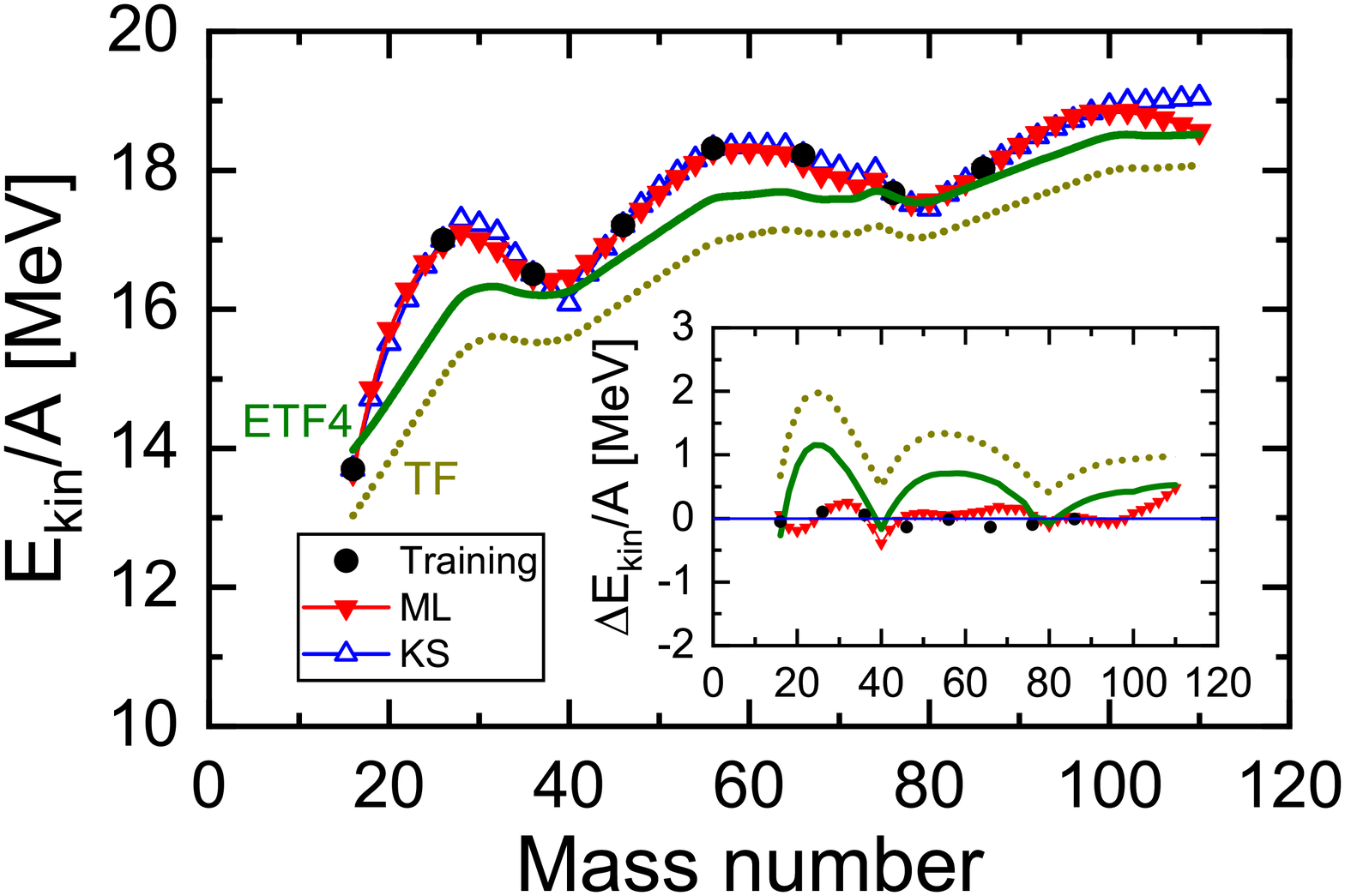}
\caption{(Color online). The total kinetic energies per nucleon of the ground states of nuclear systems with mass number from $A=16$ to $A=110$ predicted by the Kohn-Sham (KS), machine-learning (ML), Thomas-Fermi (TF), and extended Thomas-Fermi method with higher-order corrections up to order $\hbar^4$ (ETF4) approaches. The inset shows the deviations between the Kohn-Sham results and the predictions given by the ML, TF, and ETF4 approaches.}
\label{supfig1}
\end{figure}

The ML functional is constructed with the training set, and is used to predict the kinetic energies of the ground-state densities for nuclear systems with mass numbers from $A=16$ to $A=110$. The obtained results are shown in Fig.~\ref{supfig1}, comparing with the exact ones from the Kohn-Sham approach as well as the ones from the Thomas-Fermi (TF) approach and the extended Thomas-Fermi approach with higher-order corrections up to order $\hbar^4$ (ETF4).
In the TF and ETF4 approaches, the total kinetic energy is expressed by the functional of the local density,
\begin{widetext}
\begin{subequations}
  \begin{align}
    &E_{{\rm kin},\tau}^{\rm TF}=\frac{1}{2 m_N}\int d^3\bm{r}~\frac{3}{5}(3\pi^2)^{2/3}\rho_\tau^{5/3},\\
    &E_{{\rm kin},\tau}^{\rm TF4}=\frac{1}{2 m_N}\int d^3\bm{r}~\Bigg{\{}\frac{3}{5}(3\pi^2)^{2/3}\rho_\tau^{5/3}+\frac{1}{36}\frac{(\bm{\nabla}\rho_\tau)^2}{\rho_\tau}+\frac{1}{3}\Delta\rho_\tau\nonumber\\
    &\qquad\qquad +\frac{1}{6480}(3\pi^2)^{-2/3}\rho_\tau^{1/3}\left[8\left(\frac{\bm{\nabla}\rho_\tau}{\rho_\tau}\right)^4-27\left(\frac{\bm{\nabla}\rho_\tau}{\rho_\tau}\right)^2\frac{\Delta\rho_\tau}{\rho_\tau}+24\left(\frac{\Delta\rho_\tau}{\rho_\tau}\right)^2\right] \Bigg{\}},
  \end{align}
\end{subequations}
\end{widetext}
with $\tau$ represents neutron or proton~\cite{Brack1985PR}.

One can see from Fig.~\ref{supfig1} that the kinetic energies predicted by the ML density functional agree nicely with the Kohn-Sham ones.
In particular, the quantum effects corresponding to the oscillations of energies can be well reproduced by the ML functional.
The predictions given by the ML functional for systems with mass number from $A = 88$ to $A=100$ are still very good, although they are outside of the training set.
Only for systems with $A>100$, the ML predictions gradually deviate from the Kohn-Sham ones.
This clearly shows that the present ML density functional works nicely in comparison with the Kohn-Sham results.
For the TF approach, it can be seen that the kinetic energies are seriously underestimated.
Although the problem is eased in the ETF4 approach by further considering higher-order corrections,
the deviations from the Kohn-Sham results are still obvious due to the lack of sufficient quantum effects.
Similar features given by the TF and ETF4 approaches can also be found in Fig.6 of Ref.~\cite{Centelles2007AP}.

\subsection{Self-consistent calculations for the ground state}

The nuclear ground state is obtained by a variation of the energy density functional with respect to the density.
In this work, the variation of the interaction energy,
\begin{align}
  E_{\rm int}[\rho(\bm{r})] = &  \int\left( \frac{3}{8}t_{0}\rho^2 + \frac{1}{16}t_{3}\rho^{2+\gamma} \right){\rm d}^3\bm{r} \notag \\
  & + \int \frac{1}{64}(9t_1 - 5t_2 -4t_2x_2)(\nabla\rho)^2 {\rm d}^3\bm{r}, \label{Skyrme}
\end{align}
can be readily obtained as
\begin{align}
  \frac{\delta E_{\rm int}[\rho]}{\delta \rho} = & \frac{3}{4}t_{0}\rho + \frac{2+\gamma}{16}t_{3}\rho^{1+\gamma} \notag \\
  & -\frac{1}{32}(9t_{1}-5t_{2}-4t_{2}x_{2})(\Delta\rho). \label{Skyrme_fd}
\end{align}
For the kinetic energy, it reads
\begin{equation}\label{functional_derivative}
  \frac{\delta E^{\rm ML}_{\rm kin}[\rho]}{\delta \rho} =\sum_{i=1}^{m}
  \frac{w_i}{AA_i\sigma^2 }(\rho_i - \rho)K(\rho_i,\rho).
\end{equation}
Once the functional derivative is obtained, the self-consistent ground-state density can be calculated with the gradient descent method starting from a trial density $\rho^{(0)}$, in each iteration $i$, it follows
\begin{equation}\label{iteraction_density}
  \rho^{(i+1)}=\rho^{(i)}-\left.\epsilon\frac{\delta E_{\rm tot}[\rho]}{\delta\rho}\right|_{\rho=\rho^{(i)}},
\end{equation}
where $\epsilon$ is a constant between 0 and 1, whose value is determined via trading the stability and speed for the convergence.

To avoid the huge errors in the functional derivative~\eqref{functional_derivative}, principal component analysis (PCA)~\cite{Snyder2015IJQC, Abdi2010WCS} are applied.
It is found that for the aim to propose a universal machine-learning density functional for realistic nuclear systems with different mass numbers, the stability and accuracy of the iteration must be further improved.
Therefore, two new recipes, i.e., adaptive functional derivative and density renormalization, are employed.

\subsubsection{Adaptive functional derivative}

The basic idea of the PCA used here is to select $L$ major dimensions of the functional derivative in the multi-dimension density space consisting of the $M$ training densities which are closest to the given density.
It is found that the stability and accuracy of the iteration barely depend on the value of $M$, once $M$ is larger than, e.g., 50.
However, they are very sensitive to the value of $L$.
For small $L$ values, the iteration is rather stable, however, the accuracy of the results could be poor.
For large $L$ values, it is expected to be more accurate, but the iteration becomes unstable since unimportant dimensions are involved and they cause numerical noises.
To achieve a good performance for both stability and accuracy of the iteration in realistic calculations, a new recipe, i.e., adaptive functional derivative (AFD), is adopted in the present work.
For the AFD, the value of $L$ is initialized as $1$, and during the iteration, it is  automatically increased by 1 after every 50 iterations.

\begin{table}[h]
\caption{Self-consistent kinetic energies $E_{\rm kin}$ (MeV) for $^{4}$He, $^{16}$O, and $^{40}$Ca obtained by the PCA with fixed $L$ values and the present AFD, in comparison with the Kohn-Sham results.}
\begin{center}
\setlength{\tabcolsep}{1.2mm}{
\begin{tabular}{l r r r c r r}
\hline
\hline
          & $L=1$  & $L=2$  & $L=3$  & $L\geq4$ & AFD & KS \\
\hline
$^{4}$He  & ~44.41 & ~36.47 & ~36.33 & - & ~35.20       & ~35.21 \\
$^{16}$O  & 248.21 & 219.74 & 210.90 & - & 218.51       & 219.29 \\
$^{40}$Ca & 712.44 & 677.56 & 640.40 & - & 643.32       & 643.11 \\
\hline
\hline
\end{tabular}}
\end{center}
\label{sup_tab1}
\end{table}

In Table.~\ref{sup_tab1}, it lists the self-consistent kinetic energies for $^{4}$He, $^{16}$O, and $^{40}$Ca obtained by the PCA with fixed $L$ values and the present AFD, in comparison with the Kohn-Sham results.
The iteration starts from a trial density obtained from a harmonic oscillator (HO) potential
with $\hbar\omega = 150 \cdot A^{-1/3}$~MeV.
The iterations with fixed $L$ values are only stable for $L\leq3$, and the accuracy of the obtained results are rather poor in comparison with the Kohn-Sham results.
This problem can be solved with the present AFD recipe, where the $L$ values are adoptively  increased during the iteration.

\subsubsection{Density renormalization}

Another issue that causes iteration instability is the breaking of the density normalization.
This usually first occurs at the beginning of the iteration, since the functional derivatives are relatively less accurate at the trial density~\eqref{functional_derivative}.
Therefore, it could sometimes induce a slight violation of the density normalization, but this violation could be exasperated with the running iteration, so finally, no convergence can be achieved.

To solve this problem, a renormalization of the density is introduced in each iteration by requiring that $\int \tilde{\rho}(r) {\rm d}r = A$.
It is found that the density renormalization works quite well in the realistic calculations for $^{4}$He, $^{16}$O, and $^{40}$Ca with the machine-leaning density functionals obtained in the present work.
